\documentstyle[12pt]{article}
\def\ZZZ{{\hbox{ Z\kern-1.6mm Z}}}
\def\zzz{{\hbox{ z\kern-1mm z}}}

\newcommand{\RRR}{{\rm R \hskip -7pt R}}
\newcommand{\bX}{\bar X}
\newcommand{\te}{\tilde e}
\newcommand{\bx}{\bar x}
\newcommand{\bw}{\bar w}

\newcommand{\bF}{\bar F}
\newcommand{\bbF}{{\bf F}}
\newcommand{\bN}{{\bf N}}

\newcommand{\VV}{{\cal V}}

\newcommand{\AAA}{{\cal A}}

\newcommand{\FF}{{\cal F}}

\newcommand{\MM}{{\cal M}}

\newcommand{\OO}{{\cal O}}

\newcommand{\EE}{{\cal E}}
\newcommand{\ZZ}{{\cal Z}}
\newcommand{\LL}{{\cal L}}

\newcommand{\wh}{\widehat}

\newcommand{\RR}{{\cal R}}
\newcommand{\NN}{{\cal N}}

\newcommand{\SSS}{{\cal S}}

\newcommand{\be}{\begin{equation}}
\newcommand{\ee}{\end{equation}}
\newcommand{\ben}{\begin{eqnarray}\displaystyle}
\newcommand{\een}{\end{eqnarray}}
\newcommand{\refb}[1]{(\ref{#1})}
\newcommand{\p}{\partial}
\newcommand{\sectiono}[1]{\section{#1}\setcounter{equation}{0}}

\def\one{{\hbox{ 1\kern-.8mm l}}}
\def\zero{{\hbox{ 0\kern-1.5mm 0}}}

\begin{document}

{}~
{}~

\hfill\vbox{\hbox{hep-th/0603149}}\break

\vskip .6cm

\medskip

\baselineskip 20pt 

\begin{center}

{\Large \bf
Higher Derivative Corrections to
Non-supersymmetric Extremal Black Holes in $\NN=2$ 
Supergravity
}

\end{center}

\vskip .6cm
\medskip

\vspace*{4.0ex}

\centerline{\large \rm Bindusar Sahoo and Ashoke Sen}

\vspace*{4.0ex}

\centerline{\large \it Harish-Chandra Research Institute}

\centerline{\large \it  Chhatnag Road, Jhusi,
Allahabad 211019, INDIA}

\vspace*{1.0ex}

\centerline{E-mail: bindusar@mri.ernet.in, ashoke.sen@cern.ch,
sen@mri.ernet.in}

\vspace*{5.0ex}

\centerline{\bf Abstract} \bigskip

Using the entropy function formalism we compute the entropy
of extremal supersymmetric and non-supersymmetric black holes
in $\NN=2$ supergravity theories in four dimensions
with higher derivative corrections.
For supersymmetric black holes our results agree with all
previous analysis. However in some examples where the four dimensional
theory is expected to arise from the dimensional reduction of a five
dimensional theory, 
there is an apparent disagreement between our results for 
non-supersymmetric black  holes
and those
obtained by using
the five dimensional description. This indicates that
for these theories supersymmetrization of the 
curvature squared term in four dimension does not produce all the
terms which would come from the dimensional reduction of a five 
dimensional
action with curvature squared terms.

 \vfill \eject

\baselineskip 18pt

\tableofcontents

\sectiono{Introduction} \label{s1}

During the last several years study of higher derivative corrections
to the entropy of extremal supersymmetric black holes have provided
fruitful results in string 
theory\cite{9801081,9812082,9904005,9906094,
9910179,0007195,0009234,0012232,0409148,0410076,
0411255,0411272,0501014}. In many examples 
these corrections match
the appropriate corrections to the statistical entropy of the
corresponding microscopic system.
Given this success one might ask:
are there similar results for non-supersymmetric black holes? While
in general studying higher derivative corrections to the entropy of 
a generic black hole is  a difficult problem, a general method for
computing the entropy of extremal, but not necessarily
 supersymmetric
black holes was developed in 
\cite{0506177,0508042}.
This method does not
provide an explicit construction of the full 
black hole solution, but
gives a way to compute the near horizon field configuration
and entropy of an extremal black hole
with a given set of charges assuming the existence of the black
hole solution. Various other recent approaches to studying
non-supersymmetric black holes in string theory can be found
in \cite{0507096,0510024,0511117,0511215,0512138,
0602005,0603003,0511306,0601016,0601183,0602022,0602292}.

In this paper we apply the method developed in
\cite{0506177,0508042} to compute the entropy of
extremal black holes in four dimensional
$\NN=2$ supergravity theories with curvature squared type corrections.
We use fully supersymmetrized version of the action given
in \cite{9602060,9603191} and construct the entropy function for
a general extremal black hole solution following the
procedure given in \cite{0506177,0508042}. 
Extremizing
the entropy function with respect to the parameters labelling
the near horizon background gives
a set of algebraic equations for these parameters
and the value of the entropy function at the extremum gives
the entropy of the corresponding black
hole. We show that these extremization
equations admit a class of solutions
which coincide with the supersymmetric extremal black holes
studied in \cite{9801081,9812082,9904005,9906094,
9910179,0007195,0009234,9508072,9602111,
9602136,9702103,
9711053}. 
In particular we recover the supersymmetric
attractor equations of \cite{9801081,9812082,9904005,9906094,
9910179,0007195,0009234} 
in the presence of higher derivative
terms. But our method allows us to go beyond the supersymmetric
configurations and  study  higher derivative
corrections to the entropy of extremal but non-supersymmetric
black holes as well. We illustrate this by several examples.

Although no 
explicit study of higher derivative corrections
to these non-supersymmetric solutions has
been carried out before, there is a general argument due to
Kraus and Larsen\cite{0506176,0508218} 
which gives an expression for the entropy
of a class of extremal non-supersymmetric 
black holes when the four dimensional theory
comes from the dimensional reduction of a five dimensional
theory and the near horizon geometry of the black hole
solution, expressed in the five dimensional language, 
has the structure of $AdS_3\times S^2$. Unfortunately
we find that our results do not agree with the prediction
of Kraus and Larsen. The only possible explanation
for this discrepancy seems to be that adding the minimal set of
terms in the four dimensional action that is required for
supersymmetrization of the curvature squared term does not
reproduce all the terms which arise from dimensional reduction
of the curvature squared term in five dimensions.
 
The rest of the paper is organized as follows. In section \ref{s2}
we review the bosonic part of the $\NN=2$ supergravity action
with curvature squared corrections. In section \ref{s3} we
propose our ansatz for the near horizon geometry of extremal
black holes in these theories, and construct the entropy function
for these black holes. The parameters labelling the near horizon
geometry of the black hole are obtained by extremizing the entropy
function with respect to these parameters. 
In section \ref{s3a} we verify that the
entropy function constructed in section \ref{s3} is invariant
under electric-magnetic duality transformation. 
In section \ref{satt} we
show that the equations obtained by extremizing the
entropy function admit a class of solutions which
obey the well known supersymmetric attractor equations
derived in \cite{9801081,9812082,9904005,9906094,
9910179,0007195,0009234}. 
However this does not exhaust the set of
solutions of the extremization equations and there are in
general other solutions which describe the near horizon geometry
of non-supersymmetric extremal black holes. In section \ref{s4}
we use our formalism to study supersymmetric extremal black
holes in tree level
heterotic string theory compactified on $T^4\times T^2$ or
$K3\times T^2$ and reproduce the known results for the entropy
and near horizon geometry of these black holes.
Section \ref{s5} is devoted to the study of non-supersymmetric
extremal black holes in the same theory. We find an expression
for the entropy of a class of 
extremal non-supersymmetric black holes
in a power series expansion in inverse power of the magnetic charge.
In section \ref{sm} we extend our analysis of non-supersymmetric
black holes to a more general class of models describing M-theory
compactification on Calabi-Yau manifolds, and explicitly
compute the first correction to the entropy of these black holes
due to higher derivative terms. Finally in section \ref{s6} we
compare our results of sections \ref{s4}-\ref{sm} 
to the predictions of 
\cite{0506176,0508218}, assuming
that the four dimensional theory under consideration
comes from dimensional reduction of a five dimensional theory,
and that the near horizon $AdS_2\times S^2$ geometry gets lifted
to a near horizon $AdS_3\times S^2$ geometry in five 
dimensions. We find that the results do not agree. Although we
do not have a complete understanding of the origin of this
discrepancy, we suggest a possible explanation for this
apparent disagreement between the two results.

\sectiono{$\NN=2$ Supergravity Action with 
Higher Derivative Corrections} \label{s2}

The off-shell formulation of $\NN=2$ supergravity action in four
dimensions was developed in 
\cite{r1,r2,r3,r4,r5,r6,r7,9602060,9603191}. Here
we shall review this formulation following 
the notation of \cite{0007195}.
The basic bosonic fields in the theory are a set of $(N+1)$ 
complex scalar fields
$X^I$ with $0\le I\le N$
(of which one can be gauged away using a scaling 
symmetry), $(N+1)$ gauge fields $A_\mu^I$ and
the metric $g_{\mu\nu}$. Besides this the theory contains
several non-dynamical fields. These include a complex 
anti-self-dual antisymmetric
tensor field $T^-_{\mu\nu}$, a real scalar field $D$, 
a $U(1)$ gauge
field $\AAA_\mu$, an $SU(2)$ gauge field $\VV^i_{~j\mu }$, a
vector field $V_\mu$, 
a set of $SU(2)$ triplet scalar fields $Y^I_{ij}$
with $0\le I\le N$,
an SU(2) triplet scalar field 
$M_{ij}$ and an SU(2) matrix valued
scalar field $\Phi^\alpha_i$ which transforms
as a fundamental of the gauged $SU(2)$ and also a 
fundamental of a global $SU(2)$ symmetry
(see eq.(3.111) of \cite{0007195}).\footnote{Following \cite{0007195}
we 
shall be using a
non-linear multiplet as the second
compensator field in our description
of the theory.  We could also work with a description of the
theory where we use {\it e.g.}
a hypermultiplet as the second
compensator field\cite{0009234}.
The expression for the entropy function given in eq.\refb{e9} in
independent of which description we use. Also we shall be using
K-gauge condition from the beginning where we set the gauge field
associated with the dilatation symmetry of conformal supergravity
to zero\cite{0007195}.}
Here $i,j$  ($1\le i,j\le 2$) are
indices labelling the fundamental representation of 
the gauged $SU(2)$ and fields in the triplet representation
are obtained by taking the symmetric combination of a
pair of indices $i,j$. The $SU(2)$ indices are raised and lowered
by the antisymmetric tensors $\varepsilon^{ij}$ and
$\varepsilon_{ij}$ with $\varepsilon^{12}
=\varepsilon_{12}=1$. 
$\alpha$ ($1\le\alpha\le 2$)
labels the fundamental representation of the global
$SU(2)$.  We define
\ben \label{e3}
&& F^I_{\mu\nu} = \p_\mu A^I_\nu - \p_\nu A^I_\mu, 
\quad ~^*F^{I\mu\nu} = {1\over 2}
(\sqrt{-\det g})^{-1}
 \epsilon^{\mu\nu\rho\sigma} F^I_{\rho\sigma}, 
\quad F^{I\pm}_{\mu\nu}
= {1\over 2} (F^I_{\mu\nu} \pm i 
~^* F^I_{\mu\nu})
\nonumber \\
&& f_\mu^\nu = -{1\over 2} R_\mu^\nu -{1\over 4}
(D - {1\over 3}R) \delta_\mu^\nu 
+{1\over 2}\epsilon_\mu^{~\nu\rho\sigma} 
(\sqrt{-\det g})^{-1} \p_\rho \AAA_\sigma
+{ 1\over 32} T^-_{\mu\rho} T^{+\nu\rho}
\nonumber \\
&& \RR_{\mu\nu}^{~~ \rho\sigma} = 
R_{\mu\nu}^{~~\rho\sigma} +
\left(f_\mu^\rho \delta_\nu^\sigma
- f_\nu^\rho \delta_\mu^\sigma - f_\mu^\sigma \delta_\nu^\rho
+ f_\nu^\sigma \delta_\mu^\rho \right)
- {1\over 32}
(T^{-\rho\sigma} T^+_{\mu\nu} + 
T^-_{\mu\nu} T^{+\rho\sigma}) \nonumber \\
&& \RR_{~\mu\nu}^{\pm ~~ \rho\sigma} = {1\over 2}
\left( \RR_{\mu\nu}^{~~ \rho\sigma} \pm {i\over 2}
(\sqrt{-\det g})^{-1}
 \epsilon^{\rho\sigma\tau\delta} \RR_{\mu\nu\tau\delta}
\right) \nonumber \\
&& \FF^i_{~j \mu\nu} = \p_\mu \VV^i_{~j \nu} -  \p_\nu 
\VV^i_{~j \mu } + {1\over 2}\, 
\VV^i_{~k \mu} \VV^k_{~j \nu} 
- {1\over 2}\, \VV^i_{~k \nu} \VV^k_{~j \mu} \nonumber \\
&& \wh A = T^{-\mu\nu}T^-_{\mu\nu}, \quad 
\wh B_{ij} = -8 \varepsilon_{k i} \FF^k_{~j\mu\nu} 
T^{-\mu\nu} -8 \varepsilon_{k j} \FF^k_{~i\mu\nu} 
T^{-\mu\nu} \, , \nonumber \\
&& \wh F^{-\mu\nu} = -16 \RR_{\rho\sigma}^{~~\mu\nu}
T^{-\rho\sigma} \nonumber \\
&& \wh C = 64 \RR^-_{\mu\nu\rho\sigma} \RR^{-\mu\nu
\rho\sigma} \, + 32 \FF^{-i}_{~~~j\mu\nu} 
\FF^{-j\mu\nu}_{~~~i} \nonumber \\
&& \qquad \qquad - 8 T^{-\mu\nu}  
\{ (\nabla_\mu - i \AAA_\mu), (\nabla^\rho - i 
\AAA^\rho)\} T^+_{\rho\nu}
+16 T^{-\mu\nu} f_\mu^\rho T^+_{\rho\nu} \nonumber \\
\een
Here $\epsilon^{\mu\nu\rho\sigma}$ denotes the totally 
antisymmetric tensor density with 
$ \epsilon^{t r\theta\phi}=1$ and
$T^{+\mu\nu}=(T^{-\mu\nu})^*$. Anti-selfduality of $T^-_{\mu\nu}$
imposes the condition:
\be \label{econd}
T^{-\mu\nu} = -{i\over 2} \, (\sqrt{-\det g})^{-1} \, 
\epsilon^{\mu\nu\rho\sigma}\, T^-_{\rho\sigma}\, .
\ee
Note that our notation for the Riemann tensor 
$R_{\mu\nu\rho\sigma}$ differs from that
of \cite{0007195} by a $-$ sign. We take
\ben \label{eri1}
&& \Gamma^\mu_{\nu\rho} = {1\over 2} g^{\mu\sigma}
\left( \p_\nu g_{\sigma\rho} + \p_\rho g_{\sigma\nu}
- \p_\sigma g_{\nu\rho} \right) \nonumber \\
&& R^\mu_{~\nu\rho\sigma} = \p_\rho 
\Gamma^\mu_{\nu\sigma} - \p_\sigma
\Gamma^\mu_{\nu\rho} + \Gamma^\mu_{\tau\rho}
\Gamma^\tau_{\nu\sigma} - \Gamma^\mu_{\tau\sigma}
\Gamma^\tau_{\nu\rho} \nonumber \\
&& R_{\nu\sigma} = R^\mu_{~\nu\mu\sigma}, \qquad
R = g^{\nu\sigma}  R_{\nu\sigma}\, .
\een
For later use we also define
\be \label{e6a}
G^-_{I\mu\nu} = -16\pi i {\p 
\LL\over \p F^{-I\mu\nu}}\, ,
\ee
where $\sqrt{-\det g}\, \LL$ is the Lagrangian density.
In carrying out the differentiation on the right hand side of
\refb{e6a} we must treat the $F^{-I}_{\mu\nu}$ for different
$\{I,\mu,\nu\}$ as
independent variables so that under a variation of $F^I_{\mu\nu}$
\be \label{edeltal}
\delta\LL = {i\over 16\pi} \left( G^-_{I\mu\nu} \delta F^{-I\mu\nu}
- G^+_{I\mu\nu} \delta F^{+I\mu\nu}
\right)\, .
\ee

The action involving these fields is written in 
terms of the prepotential
$F(\vec X, \wh A)$, -- a meromorphic function of the complex
fields $X^I$ and the composite auxiliary field $\wh A$. 
$F$ satisfies the condition:
\be \label{e3a}
F(\lambda \vec X, \lambda^2 \wh A) = \lambda^2
F(\vec X, \wh A)\, .
\ee
We
define\footnote{Note that we are using the same symbol $F$ for the
prepotential and the gauge field strengths. This should not cause
any confusion since the index structures of these two sets
of quantities are
quite different.}
\be \label{e4}
F_I = {\p F\over \p X^I}, \quad F_{\wh A} = {\p F\over \p
\wh A}, \quad F_{IJ} ={\p^2 F\over \p X^I \p X^J},
\quad F_{\wh A I} = {\p^2 F\over \p X^I \p \wh A},
\quad F_{\wh A \wh A} =
{\p^2 F\over \p \wh A^2}\, .
\ee
In terms of the prepotential the bosonic part of
the action is given by (see eq.(3.111)
of \cite{0007195}) 
\be \label{e5}
S = \int \, d^4 x\, \sqrt{-\det g} \, \LL\, ,
\ee
where
\ben \label{e6}
8 \, \pi \, \LL &=& - {i\over 2} (X^I \bF_I - \bX^I F_I) R
+ \Bigg[ {i } (\p_\mu F_I+ i \AAA_\mu F_I )
(\p^\mu \bX^I- i\AAA^\mu \bX^I) \nonumber \\
&&
+{i\over 4} F_{IJ} (F^{-I}_{\mu\nu} -{1\over 4} \bX^I 
T^-_{\mu\nu})
(F^{-J\mu\nu} -{1\over 4} \bX^J T^{-\mu\nu}) \nonumber \\
&& +{i\over 8} \bF_I  (F^{-I}_{\mu\nu} -{1\over 4} \bX^I 
T^-_{\mu\nu}) T^{-\mu\nu} - {i\over 8} F_{IJ} Y^I_{ij}
Y^{Jij} + {i\over 32} \bF \, \wh A
\nonumber \\
&& + {i\over 2} F_{\wh A} \wh C - {i\over 8} F_{\wh A
\wh A} (\wh B_{ij} \wh B^{ij} - 2 \wh F^-_{\mu\nu}
\wh F^{-\mu\nu}) \nonumber \\
&& +{i\over 2} \wh F^-_{\mu\nu} F_{\wh A I} 
(F^{-I}_{\mu\nu} -{1\over 4} \bX^I 
T^-_{\mu\nu}) -{i\over 4} \wh B_{ij} F_{\wh A I} Y^{Iij}
\nonumber \\
&& + h.c. \Bigg]\nonumber \\
&& - i (X^I \bF_I - \bX^I F_I) 
\bigg( \nabla^\mu V_\mu - {1\over 2} V^\mu V_\mu - {1\over 4}
|M_{ij}|^2    \nonumber \\
&&  + (\p^\mu \Phi^i_\alpha + {1\over 2}\,
\VV^{i\mu}_{~~j}
\Phi^j_\alpha) (\p_\mu \Phi^\alpha_i + {1\over 2}\,
\VV^{~k}_{i\mu}
\Phi^\alpha_k)\bigg) \, . \nonumber \\
\een
Here $\Phi^j_\alpha= (\Phi^\alpha_j)^*$
and $\nabla_\mu$ denotes ordinary covariant derivative.
Furthermore, the fields are subject to the constraint
\be \label{e7}
\nabla^\mu V_\mu - {1\over 2} V^\mu V_\mu - {1\over 4}
|M_{ij}|^2  + (\p^\mu \Phi^i_\alpha + {1\over 2}\,
\VV^{i\mu}_{~j}
\Phi^j_\alpha) (\p_\mu \Phi^\alpha_i + {1\over 2}\,
\VV^{~k}_{i\mu}
\Phi^\alpha_k) - D +{1\over 3} R= 0\, .
\ee

\sectiono{Entropy Function for Extremal Black Holes} \label{s3}

In the theory described in section \ref{s2} we 
consider extremal black holes with near
horizon geometry of the 
form:\footnote{Note
that the normalization of the magnetic charge 
vector $\vec p$ used here differs from
that of \cite{0506177} by a factor of $4\pi$. 
Similarly the normalization of the electric charge vector
$\vec q$ introduced in \refb{e8} differs from
that of \cite{0506177} by a factor of $-{1\over 2}$.
These normalizations have been chosen so as to
be consistent with the ones used in
\cite{0007195}.}
\ben \label{e1}
&& ds^2 = v_1 ( - r^2 dt^2 + dr^2 / r^2) + v_2 (d\theta^2
+ \sin^2\theta d\phi^2) \nonumber \\
&& F^I_{rt} = e_I, \quad F^I_{\theta\phi}=  p^I
\, \sin\theta, \quad 
X^I = x^I, \quad
T^-_{rt}= v_1\, w \nonumber \\
&& D-{1\over 3} R = 0, \quad \AAA_\mu = 0, \quad  
\VV^{i}_{~j\mu} = 0, \quad V_\mu = 0,
\quad M_{ij}=0, \quad \Phi^\alpha_i
=\delta^\alpha_i\, , \quad Y^I_{ij}=0\, . \nonumber \\
\een
As can be seen from the action \refb{e6} and the
constraint \refb{e7},
this is a consistent truncation, respecting the symmetries of
$AdS_2\times S^2$.  In particular the equations of motion
for the fields $D$, $\AAA_\mu$, $\VV^i_{~j\mu}$, $V_\mu$,
$M_{ij}$ and $\Phi^\alpha_i$ subject to the constraint
\refb{e7}, as well as the constraint itself, are automatically
satisfied for the background \refb{e1}.
We now define the entropy function $\EE(v_1, v_2, w,
\vec x,
\vec e, \vec q, \vec  p)$ as follows\cite{0506177}
\be \label{e8}
\EE(v_1, v_2, w, \vec x,
\vec e, \vec q, \vec p) = 2\pi
\left( -{1\over 2}\,
\vec q.\vec e - \int d\theta d\phi \, 
\sqrt{-\det g} \, \LL\right)\, ,
\ee
where the $\sqrt{-\det g}\, \LL$ 
appearing on the right hand side of eq.\refb{e8}
is to be evaluated for the background \refb{e1} and the integral
over $\theta$, $\phi$ is to be evaluated at fixed $r$, $t$. 
For an extremal black hole carrying electric charge 
vector $\vec q$ and magnetic charge vector $\vec  p$
the 
parameters $\vec x$, $\vec e$, $v_1$, $v_2$, $w$ labelling
the near
horizon geometry are
obtained by extremizing the function $\EE$ with respect to $e^I$,
$x^I$, $v_1$, $v_2$ and $w$: 
\be \label{extreme}
{\p\EE\over \p v_i}=0, \qquad {\p\EE\over \p x^I}=0, \qquad
{\p\EE\over \p w}=0, \qquad {\p\EE\over \p e^I}=0\, ,
\ee
and the entropy 
associated with the black hole
is given by the value of the function $\EE$ at the 
extremum\cite{0506177,0508042}:
\be \label{eblack}
S_{BH} = \EE\, .
\ee
Comparing \refb{e6a} with the equation
obtained by extremizing \refb{e8} with respect to $e^I$:
\be \label{eqi1}
q_I = -2\,
{\p\over \p e^I} \int d\theta d\phi \, \sqrt{-\det g} \, \LL\, ,
\ee
and using the definition of $e^I$ given in \refb{e1},
we can show that near the horizon
\be \label{eqi2}
G_{I\theta\phi} =  q_I \, \sin\theta\, .
\ee

We can now calculate the various quantities defined in eq.\refb{e3}
for the background described in \refb{e1}. In particular  
we get
\ben \label{eyy0}
&& f^r_r = f^t_t = {1\over 2} v_1^{-1} -{1\over 32} w\bar w\, ,
\qquad f^\theta_\theta = f^\phi_\phi = -{1\over 2} v_2^{-1}
+{1\over 32} w\bar w \nonumber \\
&& f^\mu_\nu=0 \quad \hbox{otherwise,} \nonumber \\
&& \RR_{m\alpha}^{~~~n\beta} = - \RR_{\alpha m}^{~~~n\beta}
= - \RR_{m\alpha}^{~~~\beta n} = \RR_{\alpha m}^{~~~\beta n}
= {1\over 2} (v_1^{-1} - v_2^{-1}) \delta_m^n
\delta_\alpha^\beta  \nonumber \\
&& \qquad \quad \hbox{for} \quad
\alpha,\beta=r,t, \quad m,n=\theta,\phi \nonumber \\
&& \RR_{\mu\nu}^{~~~\rho\sigma} =0 \quad \hbox{otherwise,} 
\een
\ben \label{eyy1}
&& \wh A = - 4 w^2, \qquad \wh B_{ij}=0, \qquad
\wh F^-_{\mu\nu} = 0, \nonumber \\ &&
\wh C = 16 w\bw \left(-v_1^{-1} - v_2^{-1} + {1\over 8} 
w\bw \right) + 128 (v_1^{-1} - v_2^{-1})^2\, .
\een
For the action given in \refb{e6}, a straightforward calculation 
now gives:
\ben \label{e9}
\EE &=& -\pi q_I e^I 
-\pi v_1 v_2 \Bigg[ i (v_1^{-1} - v_2^{-1})
(x^I \bF_I - \bx^I F_I) \nonumber \\
&& -\Big\{
{i\over 4} v_1^{-2} F_{IJ}
(e^I - i v_1 v_2^{-1} p^I - {1\over 2} \bx^I v_1 w) 
(e^J - i v_1 v_2^{-1} p^J - {1\over 2} \bx^J v_1 w)
 + h.c. \Big\} 
 \nonumber \\
 &&  -\Big\{ {i\over 4} v_1^{-1} w \bF_I (e^I - 
 i v_1 v_2^{-1} p^I - {1\over 2} \bx^I v_1 w) + h.c.\Big\}
 \nonumber \\
 && +\Big\{{i\over 8} \bw^2 F + h.c.\Big\}
 + 8 \, i \,
 \bw w\Big( - v_1^{-1} - v_2^{-1} + {1\over 8} \bw w
 \Big)
 \Big(F_{\wh A} - \bF_{\wh A} \Big)  \nonumber \\
 && + 64\, i \, (v_1^{-1} - v_2^{-1})^2
  \Big(F_{\wh A} - \bF_{\wh A} \Big) \Bigg]\nonumber \\
  &\equiv& -\pi q_I e^I - \pi \, g(v_1, v_2, w, \vec x, 
  \vec e, \vec p)
  \, .
\een
The entropy function defined here has a scale invariance
\be \label{e10}
x^I \to \lambda x^I, \quad v_i \to \lambda^{-1}\bar\lambda^{-1}
v_i, \quad e^I \to e^I. \quad w\to \lambda w, \quad
q_I\to q_I, \quad p^I\to p^I\, .
\ee
This descends from the invariance of the lagrangian density
\refb{e6} under local scale transformation,
and is usually eliminated by 
using some gauge fixing
condition. We shall however find it convenient to work 
with the gauge
invariant equations of motion obtained by extremizing \refb{e9}
with respect to $v_1$, $v_2$, $w$, $\vec x$ and $\vec e$. 

Since \refb{e9} is quadratic in the
electric field variables $e^I$ we can explicitly eliminate
them by solving their equations of motion to express the
entropy function as a function of the other variables.
A tedious but straightforward algebra shows that after eliminating
the variables $e^I$ the entropy function reduces to:
\ben \label{eted1}
\EE(v_1, v_2, w, \vec x, \vec q, \vec p)
&=& \pi \Bigg[ i (v_1-v_2) (x^I \bar F_I - \bar x^I F_I) \nonumber \\
&&
+ v_1 v_2^{-1}
\pmatrix{p^I & q_I} \, \pmatrix{({\bf \bar 
F N^{-1} F})_{IJ} &  
-({\bf \bar F N^{-1}})_I^{~J} \cr - 
({\bf N^{-1} F})^I_{~J} & 
(\bN^{-1})^{IJ}} \, \pmatrix{p^J \cr q_J}\nonumber \\
&& -{i\over 2} v_1 \left\{
w \pmatrix{p^I & q_I} \, \pmatrix{({\bf \bar 
F N^{-1} F})_{IJ} &  
-({\bf \bar F N^{-1}})_I^{~J} \cr - 
({\bf N^{-1} F})^I_{~J} & 
(\bN^{-1})^{IJ}}   \, \pmatrix{\bar x^J \cr \bar F_J} - h.c.
\right\} \nonumber \\
&& + 8 v_1 v_2 \bar w^3 w^3 (\bN^{-1})^{IJ}
  \bar F_{\wh A I}  F_{\wh A J} \nonumber \\
&& + \left\{
4 i v_1 v_2 \bar w^2 w^4   
\left( F_{\wh A\wh A} + i\,
(\bN^{-1})^{IJ} F_{\wh A I} F_{\wh A J}
\right)
+ h.c.\right\}
\nonumber \\
&& +{i\over 8} v_1 v_2 w \bar w (x^I \bar F_I - \bar x^I F_I)
+ 8 i w \bar w (v_1 + v_2 - {1\over 16} v_1 v_2
w \bar w)
(F_{\wh A} - \bar F_{\wh A}) \nonumber \\
&& - 64\, i \, v_1 v_2\,
\left(v_1^{-1} - v_2^{-1} \right)^2
  \Big(F_{\wh A} - \bF_{\wh A} \Big)
\Bigg]\, ,  
\een
where
\be \label{eted2}
N_{IJ} = i \left( \bar F_{IJ} - F_{IJ} \right)\, ,
\ee
and $\bN$, $\bbF$ denote matrices with matrix elements $N_{IJ}$
and $F_{IJ}$ respectively.
Note that by an abuse of notation we have continued to
use the symbol
$\EE$ to denote the entropy function even
after elimination of the
variables $e^I$.  In arriving at \refb{eted1} we have used the
relations
\be \label{euler}
x^I F_I + 2\wh A F_{\wh A} = 2 \, F, \qquad 
x^I F_{IJ} + 2\wh A F_{J\wh A} = F_J, \qquad
x^I F_{I\wh A}+ 2\wh A F_{\wh A \wh A} = 0\, ,
\ee
which follow from \refb{e3a}.

\sectiono{Symplectic Invariance of the Entropy Function}
\label{s3a}

As has been discussed in \cite{9801081,9812082,9904005,9906094,
9910179,0007195,0009234}, the equations of
motion derived from the Lagrangian density \refb{e6} retain
their form under a symplectic transformation:
\be \label{esymp1}
\pmatrix{\check X^I \cr \check F_J}
= \pmatrix{U^I_{~K} & Z^{IL} \cr W_{JK} & V_J^{~L}}
\pmatrix{X^K\cr F_L}\, ,  \qquad
\pmatrix{\check F^{\pm I}_{\mu\nu} \cr \check 
G^{\pm}_{J\mu\nu}}
= \pmatrix{U^I_{~K} & Z^{IL} \cr W_{JK} & V_J^{~L}}
\pmatrix{F^{\pm K}_{\mu\nu}\cr G^{\pm}_{L\mu\nu}}\, , 
\ee
with all other fields, including the metric $g_{\mu\nu}$ 
and the
auxiliary field $T^-_{\mu\nu}$, remaining invariant. 
Here $U$, $Z$, $W$ and $V$ are each $(N+1)\times
(N+1)$ matrix, satisfying the conditions
\be \label{esymp1a}
U^T W - W^T U =0, \qquad Z^T V - V^T Z =0, \qquad
U^T V - W^T Z = {\bf 1}\, ,
\ee
so that $\pmatrix{U & Z\cr W & V}$ is a symplectic
matrix.
Eq.\refb{esymp1}
not only tells us how the fundamental
fields $X^I$ and $F^I_{\mu\nu}$ transform under this 
transformation, but also implicitly tells us how the prepotential
$F$ transforms to a new prepotential $\check F$ (so that
$\check F_I = \p \check F / \p \check X^I$). Since in general
$\check F$ and $F$ have different functional forms, the
transformation \refb{esymp1} is not a symmetry. In special
cases where $\check F$ and $F$ have the same form, the
symplectic transformations generate (continuous) duality
symmetries of the classical theory.

{}From \refb{e1}, \refb{eqi2} it follows that under a 
symplectic transformation the parameters labelling the
near horizon geometry of a black hole transform as
\ben \label{esymp2}
&& \pmatrix{\check x^I \cr \check F_J}
= \pmatrix{U^I_{~K} & Z^{IL} \cr W_{JK} & V_J^{~L}}
\pmatrix{x^K\cr F_L}\, , \qquad
\pmatrix{\check p^I \cr \check q_J}
= \pmatrix{U^I_{~K} & Z^{IL} \cr W_{JK} & V_J^{~L}}
\pmatrix{p^K\cr q_L}\, , \nonumber \\ \cr
&&
 \check v_1 = v_1, \qquad \check v_2 = v_2, \qquad
\check w = w\, .
\een

We shall now  verify that the entropy function \refb{eted1}
is invariant under the symplectic transformation. 
Using the relations 
(see {\it e.g.} eqs.(3.88), 
(3.97)-(3.99) of \cite{0007195})
\ben \label{eted4}
&&
\check F_{\wh A} = F_{\wh A}\, , \qquad {\bf \check F}
= (V {\bf F} + W) (U + Z {\bf F})^{-1}\, , \qquad 
\check \bN^{-1} = \bar\SSS \bN^{-1} 
\SSS^T = \SSS \bN^{-1} \bar \SSS^T,
\nonumber \\
&& \check F_{\wh A I} = F_{\wh A J}(\SSS^{-1})^J_{~I},
\qquad \check F_{\wh A\wh A} = F_{\wh A\wh A}
- F_{\wh A I} F_{\wh A J} \ZZ^{IJ}
\een
where
\be \label{eted5}
\SSS^I_{~J} = U^I_{~J} + Z^{IK} F_{KJ}\, ,
\qquad \ZZ^{IK} = (\SSS^{-1})^I_{~J} Z^{JK}\, ,
\ee
it is easy to check that
\be \label{eted4b}
\check F_{\wh A\wh A} +i (\check \bN^{-1})^{IJ} 
\check F_{\wh A I} \check F_{\wh A J}
= F_{\wh A\wh A} + i(\bN^{-1})^{IJ} F_{\wh A I} F_{\wh A J}\, ,
\ee
and
\be \label{eted3} 
\pmatrix{\bf \check{\bar F} \check N^{-1} \check
F  &  
-\bf \check{\bar F} \check N^{-1}  \cr - 
\bf \check N^{-1} \check F  & 
\bf \check N^{-1} } = \pmatrix{ V & - W \cr - Z & U}
\pmatrix{ \bf \bar F N^{-1} F  &  
-\bf \bar F N^{-1}  \cr - \bf N^{-1} F  & 
\bf N^{-1} } \pmatrix{V^T & - Z^T \cr - W^T & U^T}\, .
\ee
Using \refb{esymp1}-\refb{eted3}
 it is straightforward to verify that the entropy
function given in \refb{eted1} is invariant under a 
sympletic transformation. This is in accordance with the
general result on duality invariance of the entropy
function discussed in \cite{0508042}.

\sectiono{Supersymmetric Attractors} \label{satt}

It can be easily seen that the  extremization
equations \refb{extreme} can be 
satisfied by setting 
\be \label{e11b} 
v_1 = v_2 = {16\over \bw w} \, ,
\ee
\be \label{e11}
e^I - i v_1 v_2^{-1} p^I - {1\over 2} \bx^I v_1 w = 0
\ee
\be \label{e11a}
(\bw^{-1}\bF_I - w^{-1} F_I) = -{i\over 4} \, q_I \, .
\ee
Taking the real and imaginary parts of eq.\refb{e11} gives
\be \label{e11c}
e^I = 4 (\bw^{-1}\bx^I + w^{-1} x^I) \, ,
\ee
and
\be \label{e11d}
(\bw^{-1}\bx^I - w^{-1} x^I) = - {1\over 4} \, i\, p^I\, .
\ee 
The black hole entropy computed using Wald's 
formalism\cite{9307038,9312023,9403028,9502009}
is equal to the entropy 
function evaluated for this background\cite{0506177} and
is given by
\be \label{e12}
S_{BH} = 2\pi\left[
-{1\over 2}\, \vec q \cdot \vec e -16 \, i \, (w^{-2} F - 
\bw^{-2} \bF) \right]
\, .
\ee
If we choose $w$=constant gauge
(which corresponds to $\wh A$=$-4 w^2$=constant), 
then eqs.\refb{e11b}-\refb{e11d}
describe the usual attractor equations for the near horizon
geometry of extremal black holes, and \refb{e12}
gives the expression for the
entropy of these black holes as written down in 
\cite{0405146}. For example
\refb{e12} shows that in the gauge  $w=$real constant,
the Legendre transform of the black hole entropy
with respect
to the electric charges $q_I$ is proportional to the
imaginary part of the prepotential $F$. Furthermore
eqs.\refb{e11b}, \refb{e11} shows that the argument
$x^I$ of the prepotential is proportional to $e^I+ip^I$, \i.e. its
real part is the variable conjugate to the electric charge $q_I$ and
its imaginary part is the magnetic charge $p^I$. These are some of
the  observations made in \cite{0405146}.

Note that the attractor equations \refb{e11b}-\refb{e11d}
provide sufficient but not
necessary conditions for extremizing the entropy function.
In section \ref{s5} we shall find near horizon configurations
which extremize the entropy function but do not satisfy
eqs.\refb{e11b}-\refb{e11d}.

\sectiono{Supersymmetric Black Holes in the STU Model} \label{s4}

Let us now restrict our attention to a specific theory with three
vector multiplets and a prepotential
\be \label{e13}
F(X^0, X^1, X^2, X^3, \wh A) = - {X^1  X^2 X^3\over X^0}
- C \wh A \, {X^1\over  X^0}\, .
\ee
For $C=1/64$ this describes a subsector of the low energy effective
action for tree level heterotic string
theory on $T^4\times T^2$ or $K3\times T^2$, with the identification
\be \label{e14}
{X^1\over X^0} = i S\, , \quad {X^2\over X^0} = i T\, , \quad
{X^3\over X^0} = i U\, ,
\ee
where $S$, $T$ and $U$ denote the usual axion-dilaton field,
the Kahler modulus of $T^2$ and the complex structure modulus
of $T^2$ respectively. The corresponding gauge fields $A_\mu^0$,
$\ldots$ $A_\mu^3$ may be identified  as the components of the 
metric and the rank two anti-symmetric tensor field with one index
along one of the directions of $T^2$ and the other index along a
non-compact direction.\footnote{In order to make this
identification we need to dualize the gauge field $\AAA_\mu^1$.
This is reflected in the relation \refb{e15} between the charges
$(\vec q, \vec p)$ in this theory and the charges $(\vec Q, \vec P)$
in heterotic string compactification.}

For the choice of the prepotential given in
\refb{e13} the equations of motion
derived from the lagrangian density
\refb{e6} are invariant under the $SO(2,2)=SL(2,R)\times SL(2,R)$
T-duality transformation:
\ben \label{ez1}
X^0\to c X^3 + d X^0\, ,   \quad
X^1\to -c F_2 + d X^1\, , \nonumber \\
X^2\to -c F_1 + d X^2\, , \quad
X^3\to a X^3 + b X^0\, , \nonumber \\
F_0\to a F_0 - b F_3\, , \quad
F_1\to a F_1 - b X^2\, , \nonumber \\
F_2\to a F_2 - b X^1\, , \quad
F_3\to - c F_0 + d F_3\, , \nonumber \\
F^{-3}_{\mu\nu} \to a \, F^{-3}_{\mu\nu} +
b \, F^{-0}_{\mu\nu}, \quad F^{-0}_{\mu\nu}\to
c \, F^{-3}_{\mu\nu} + d \, F^{-0}_{\mu\nu}, \nonumber \\
G^-_{1\mu\nu} \to a \, G^-_{1\mu\nu}
- b \, F^{-2}_{\mu\nu}\, , \quad
F^{-2}_{\mu\nu} \to -c \, G^-_{1\mu\nu}+ d
\, F^{-2}_{\mu\nu}
\nonumber \\
G^-_{3\mu\nu} \to d \, G^-_{3\mu\nu} -
c \, G^-_{0\mu\nu}, \quad G^-_{0\mu\nu}\to
-b \, G^-_{3\mu\nu} + a \, G^-_{0\mu\nu}, \nonumber \\
F^{-1}_{\mu\nu} \to d \, F^{-1}_{\mu\nu}
- c  \, G^-_{2\mu\nu}\, , \quad
G^-_{2\mu\nu} \to -b \, F^{-1}_{\mu\nu}+ a
\, G^-_{2\mu\nu}
\een
and 
\ben \label{ez1a}
X^0\to r X^2 + s X^0\, ,   \quad
X^1\to -r F_3 + s X^1\, , \nonumber \\
X^3\to -r F_1 + s X^3\, , \quad
X^2\to k X^2 + l X^0\, , \nonumber \\
F_0\to k F_0 - l F_2\, , \quad
F_1\to k F_1 - l X^3\, , \nonumber \\
F_3\to k F_3 - l X^1\, , \quad
F_2\to - r F_0 + s F_2\, , \nonumber \\
F^{-2}_{\mu\nu} \to k \, F^{-2}_{\mu\nu} +
l \, F^{-0}_{\mu\nu}, \quad F^{-0}_{\mu\nu}\to
r \, F^{-2}_{\mu\nu} + s \, F^{-0}_{\mu\nu}, \nonumber \\
G^-_{1\mu\nu} \to k \, G^-_{1\mu\nu}
- l \, F^{-3}_{\mu\nu}\, , \quad
F^{-3}_{\mu\nu} \to -r \, G^-_{1\mu\nu}+ s
\, F^{-3}_{\mu\nu}
\nonumber \\
G^-_{2\mu\nu} \to s \, G^-_{2\mu\nu} -
r \, G^-_{0\mu\nu}, \quad G^-_{0\mu\nu}\to
-l \, G^-_{2\mu\nu} + k \, G^-_{0\mu\nu}, \nonumber \\
F^{-1}_{\mu\nu} \to s \, F^{-1}_{\mu\nu}
- r  \, G^-_{3\mu\nu}\, , \quad
G^-_{3\mu\nu} \to -l \, F^{-1}_{\mu\nu}+ k
\, G^-_{3\mu\nu}\, ,
\een
where
\be \label{esltr}
ad - bc = 1, \qquad ks - lr = 1, \qquad
a,b,c,d,k,l,r,s\in \RRR\, .
\ee
These are special cases of the symplectic transformations discussed
in section \ref{s3a} for which 
$\check F$ has the same functional form as $F$.

We now define:
\ben \label{e15}
&& Q_1 =   q_2, \quad Q_2 = - p^1, \quad Q_3 =  q_3, 
\quad Q_4 =  q_0\, , \nonumber \\
&& P_1 = p^3, \quad P_2 = p^0, \quad P_3 = p^2, \quad
P_4 =   q_1\, ,
\een
\be \label{e15b}
Q^2 = 2(Q_1 Q_3 + Q_2 Q_4), \quad P^2 = 2 (P_1 P_3 + P_2 P_4),
\quad Q\cdot P = (Q_1 P_3 + Q_3 P_1 + Q_2 P_4 + Q_4 P_2)
\, ,
\ee
where $p^I$ and
$q_I$ have been defined in eqs.\refb{e1} and \refb{eqi2}.
{}From \refb{ez1}-\refb{e15}
it follows that the duality 
transformations act on $\vec P$, $\vec Q$ as $SO(2,2)$
transformations:
\be \label{sott}
\pmatrix{Q_1 \cr Q_2 \cr Q_3\cr Q_4} \to 
\Omega \, \pmatrix{Q_1 \cr Q_2 \cr Q_3\cr 
Q_4}, \qquad
\pmatrix{P_1 \cr P_2 \cr P_3\cr P_4} \to 
\Omega \, \pmatrix{P_1 \cr P_2 \cr P_3\cr P_4}\, ,
\ee
where $\Omega$ is the $SO(2,2)$ matrix
\be \label{eomega}
\Omega = \pmatrix{s & 0 & 0 & -r\cr 0 & s & r & 0 \cr
0 & l & k & 0\cr -l & 0 & 0 & k}\, 
\pmatrix{a & b & 0 & 0\cr c & d & 0 & 0\cr 0 & 0 & d & -c\cr
0 & 0 & -b & a} \, .
\ee
$Q^2$, $P^2$ and $Q\cdot P$ defined in \refb{e15b}
are the three 
independent duality invariant combinations which can be formed out of
$\vec Q$ and $\vec P$. Thus
the black hole entropy
depends only on these combinations.

Supersymmetric black holes in this theory have been 
analyzed in
detail in \cite{9801081,9812082,9904005,9906094,
9910179,0007195,0009234}. 
These  black holes exist for $P^2>0$, $(Q\cdot P)^2 
< Q^2 P^2$ and the entropy associated with these black holes can
be obtained by extremizing the entropy function with respect to
various near horizon parameters, and plugging them into 
\refb{e12}. 
The solution satisfies the supersymmetric attractor equations
given in \refb{e11b}-\refb{e11d}.
Due to the duality symmetry \refb{sott} we can choose to work
in a special frame in which $P_2=0$, \i.e. $p^0=0$. By solving
the attractor equations \refb{e11b}-\refb{e11d} and using the 
definitions \refb{e15}, \refb{e15b} we get, in the $w=1$ gauge,
 \ben \label{ew1} &&
x^0 = -{1\over 8} Q_2 \sqrt{P^2 (P^2+512 C) \over
 P^2 Q^2 - (P\cdot Q)^2} \nonumber \\
&& {x^1\over x^0} = -{P\cdot Q\over P^2} + i \sqrt{P^2 Q^2 -
(P\cdot Q)^2 \over P^2 (P^2+512 C)} \nonumber \\
&&{x^2\over x^0} = -{1\over 2 Q_2 P_1} (Q_2 P_4 + Q_1 P_3 - P_1
Q_3) - i {P_3\over Q_2}\sqrt{P^2
Q^2 - (P\cdot Q)^2 \over P^2 (P^2+512 C)} \nonumber \\
&&{x^3\over x^0} = -{1\over 2 Q_2 P_3} (Q_2 P_4 - Q_1 P_3 + P_1
Q_3) - i {P_1\over Q_2}\sqrt{P^2 Q^2 - (P\cdot Q)^2 \over P^2
(P^2+512 C)} \nonumber \\
&& v_1 = v_2 =16, \qquad e^I = 8 \, Re(x^I) \quad \hbox{for} \quad
0\le I\le 3, \qquad w=1\, .
 \een
 The entropy associated with these black holes is given by
 \be \label{ew2}
 S_{BH} = \pi \sqrt{P^2 Q^2 - (P\cdot Q)^2} \sqrt{1 + {512 C\over
 P^2}}\, .
 \ee
This result was derived in \cite{9906094} and reviewed in 
eq.(6.64) of 
\cite{0007195}.

The solution simplifies for a 
specific class of black holes for which $P\cdot Q=0$.
In this case we can get
supersymmetric black holes if $Q^2>0$, $P^2>0$. 
A representative element satisfying this condition is
\be \label{e23aa}
P_1 = P_3 = P_0, \qquad Q_2 = Q_4 = -Q_0, \qquad 
P_2 = P_4 =  Q_1 = Q_3 = 0, \qquad
Q_0, P_0>0\, ,
 \ee
 with $P_0>0$, $Q_0>0$.
In this case
\be \label{eex1}
Q^2 = 2 Q_0^2, \qquad P^2 = 2 P_0^2\, ,
\ee
and, according to \refb{e15},  
 \be \label{exx4}
 p^1 = Q_0, \quad p^2 = P_0, \quad p^3 = P_0, \quad q_0 =
 -  Q_0\, ,
 \ee
 with all other charges zero.
Eqs.\refb{ew1}, \refb{ew2} now reduce to:
\ben \label{eex2}
&& x^0 = {1\over 8}\,  \sqrt{P_0^2 + 256 C}, 
\quad x^1 = {i\over 8} Q_0, \quad
x^2 = {i\over 8} P_0, \quad x^3
= {i\over 8} P_0, \nonumber \\
&& e^0 =   \sqrt{P_0^2 + 256 C},  \quad e^1 = e^2 = e^3 = 0\, , 
\quad w =1\, ,
\nonumber \\
&& v_1 = 16, \quad v_2 = 16\, ,
\een
and
\be \label{ex2}
S_{BH} = 2\pi Q_0 \sqrt{P_0^2 + 256 C }
= \pi \, \sqrt{Q^2 \, P^2 } \, 
\sqrt{1 + 512 {C \over P^2}}\, .
\ee

\sectiono{Non-supersymmetric Extremal Black Holes 
in the STU Model} \label{s5}

For $C=0$ the theory described in section \ref{s4}
also contains extremal non-supersymmetric black
holes 
for\cite{9702103,0507096,0510024,0511117,0511215,0512138,
0602005,0603003}
\be \label{ens1}
Q^2 P^2 < (Q\cdot P)^2\, .
\ee
As before, 
the entropy of these black 
holes can be obtained by extremizing the
entropy function. 
Due to the duality symmetries given in 
\refb{sott}, \refb{eomega}, we can
simplify the calculation by choosing a representative 
$\vec Q$, $\vec P$ satisfying
\be \label{esat1}
P_2 = P_4 = Q_2 = Q_4 = 0\, ,
\ee
and then rewriting the final result in a duality invariant form.
It turns out that in this case the resulting entropy function,
after elimination of the auxiliary variable $w$, and the electric
field variables $e^I$, has
a $Z_2$ symmetry which allows us to set\footnote{Physically this
corresponds to a solution in heterotic string theory on $T^4\times T^2$
or $K3\times T^2$ where the electric and magnetic charges
associated with only one of the two circles of $T^2$ are present.
Thus $T^4\times S^1$ or $K3\times S^1$ part factorizes from the black
hole geometry.}
\be \label{esat2}
Im(T)=Im(U)=0\, .
\ee
The final result for the entropy function 
after extremization is
\be \label{ens2}
S_{BH} = \pi\sqrt{(Q\cdot P)^2 - Q^2 P^2}\, .
\ee

For simplicity we shall focus our attention on a
special class of these black holes for which
\be \label{e22}
Q\cdot P=0, \qquad P^2 > 0, \qquad Q^2 < 0\, .
\ee
In this case instead of using the configuration \refb{esat1} we shall
use
a representative element
\be \label{e23}
-Q_2 = Q_4 = Q_0, \quad P_1 = P_3 = P_0, \quad 
Q_1 = Q_3 = P_2 = P_4 = 0, \qquad Q_0,P_0>0\, ,
\ee
so that we have
\be \label{e23a}
P^2 = 2 P_0^2, \qquad Q^2 = - 2 Q_0^2\, ,
\ee
and, according to \refb{e15},
 \be \label{ez5}
 p^1 = Q_0, \quad p^2 = P_0, \quad p^3 = P_0, \quad q_0 =
 Q_0\, .
 \ee
Note that the charge assignment \refb{e23} differs from that of
\refb{e23aa} by simple reversal of the sign of $Q_4$, \i.e. of
$q_0$. 
For $C=0$ the bosonic part of the
action,  after elimination of the auxiliary
field $w$, has a $Z_2$ symmetry that allows us to
relate the black hole solutions for the charge configurations
\refb{e23aa} and \refb{e23} 
by simple reversl of the sign of $e^0$.
Thus the near horizon geometry for the 
black hole solution corresponding to the charges given in \refb{e23}
is obtained from \refb{eex2} by setting $C=0$, changing the sign
of $e^0$, and finally determining $w$ by solving its equation
of motion. This gives
\ben \label{eexy2}
&& x^0 = {1\over 8}\, P_0 , \quad x^1 = {i\over 8}   Q_0, 
\quad
x^2 = {i\over 8} P_0, \quad x^3
= {i\over 8} P_0, \nonumber \\
&& e^0 = -P_0, \quad e^1 = e^2 = e^3 = 0, 
\quad w ={1\over 2}\, ,
\nonumber \\
&& v_1 = 16, \quad v_2 = 16\, .
\een
One can verify explicitly that this configuration extremizes the
entropy function \refb{e9} for
$F(\vec X, \wh A)=-X^1X^2X^3/X^0$.
The corresponding entropy is
\be \label{eexy3}
S_{BH} = 2\pi \, Q_0 \, P_0\, ,
\ee
in accordance with the general result \refb{ens2}.
Our goal will be to analyze higher derivative corrections
to the near horizon geometry of these
non-supersymmetric black holes by keeping $C\ne 0$.
In order to  do so, it will be  convenient to choose the gauge
\be \label{e23b}
w = {1\over 2}\, 
\ee
so that the leading order solution given in \refb{eexy2} already
satisfies the gauge condition.
In this gauge the entropy function evaluated for 
$\vec P$, $\vec Q$ of the form
given in \refb{e23} can be shown to be invariant under
the transformation
\be \label{etrs}
x^0 \to (x^0)^*, \qquad x^i\to -(x^i)^* \quad \hbox{for} \quad
1\le i \le 3, \qquad e^i \to - e^i
\quad \hbox{for} \quad
1\le i \le 3\, .
\ee
Thus we can look for a solution to the extremization equation
within the subspace which is invariant under the transformation
\refb{etrs}, \i.e. we take
\be \label{etrs1}
x^0 = (x^0)^*, \qquad x^i =- (x^i)^* \quad \hbox{for} \quad
1\le i \le 3, \qquad e^i = 0 \quad \hbox{for} \quad
1\le i \le 3\, .
\ee
It will be convenient to
introduce rescaled real varibles $y^0$, $y^1$, $y^2$, $y^3$,
$\te^0$ through
\be \label{e23d}
x^0 =P_0 y^0, \quad
x^1 = i Q_0 y^1, \quad x^2 = i P_0 y^2, \quad 
x^3 = i P_0 y^3, \quad e^0 = P_0 \te^0\, .
\ee
Substituting \refb{ez5}, \refb{e23b}-\refb{e23d} into \refb{e9} we get
\ben \label{e27}
\EE &=& \pi Q_0 P_0 \Bigg[ - \te^0 - {v_1\over v_2} 
{y^1+y^2+y^3\over y^0}
  + \left\{ {v_1 \over y^0}- {\te^0  
\over (y^0)^2}\right\}  (y^1 y^2+ y^2 y^3 + y^1 y^3) 
\nonumber \\
&&
+\left\{ - {(\te^0)^2 v_2 
\over v_1 (y^0)^3} + {\te^0 v_2  \over (y^0)^2}
+ { 8 (v_2 - v_1) \over y^0}-{v_1 v_2
\over 2 y^0} \right\} y^1 y^2 y^3 \nonumber \\
&&   + {C\over P_0^2} \Bigg\{
- {\te^0\over (y^0)^2} + {v_1\over 2 y^0} -{(\te^0)^2 v_2 y^1
\over v_1 (y^0)^3} +{\te^0 v_2 y^1\over 2 (y^0)^2}
+ {8 v_2 y^1\over y^0} -{3 v_1 v_2 y^1
\over 16 y^0}\nonumber \\
&&  - {128y^1\over y^0} \left({v_1\over v_2}
-{v_2\over v_1}\right)^2
\Bigg\}
\Bigg] \nonumber \\
\een
Note that $Q_0P_0={1\over 2}
\sqrt{-Q^2 P^2}$ 
appears as an overall factor in the above
expression, and the rest of the expression is a function of the
combination $C/P^2=C/(2 P_0^2)$. 
Thus the black hole entropy, obtained by 
extremizing \refb{e27} with respect to 
$v_1$, $v_2$, $\te^0$,  $y^0$, $y^1$, $y^2$
and $y^3$, must be of the form:
\be \label{e28}
S_{BH} =   
\sqrt{-Q^2 P^2} \, f\left({C\over P^2}\right)\, ,
\ee
for some function $f(u)$. We shall try to analyze $f(u)$
as a power series expansion in $u$. The leading contribution,
which corresponds to setting the term involving $C/P^2$ in
\refb{e27} to zero, is given by
\be \label{e28a}
f(0) = \pi \, .
\ee
The corresponding values of $v_1$, $v_2$, $\te^0$ and $y^I$ are
given by
\be  \label{e29}
v_1 = 16, \quad v_2 = 16, \quad \te^0 = -1, \quad 
y^0 = 1/8, \quad y^1 = 1/8, \quad y^2 = 1/8, \quad y^3 = 1/8\, .
\ee
These results are in agreement with \refb{eexy2}, \refb{eexy3}.

The order $u$ term in $f(u)$ can be obtained by 
evaluating the order $C/P^2$ term in \refb{e27} in the
background \refb{e29}. This gives
\be \label{e30}
f(u) = \pi (1 + 80\, u + \OO(u^2))\, .
\ee
In order to determine the higher order corrections to $f(u)$
we need to
solve the extremization equations iteratively as a
power series in $u$. The result for the first few terms is
\ben \label{e31}
f(u) &=& \pi( 1 + 80 u  - 3712 u^2  - 243712 u^3 
- 18325504 u^4  - 9538502656 u^5 \nonumber \\
&&
+   7416509890560   u^6   + 1770853956059136  u^7
 + 32680138894213120  u^8 \nonumber \\
&&  -
 194861291843407052800 u^9 
  - 115321933038468181524480 u^{10}
+ \ldots 
) \nonumber \\
\een

\sectiono{Black Holes in M-theory on  Calabi-Yau
Manifolds} \label{sm}

In this section we shall repeat the analysis of the 
previous sections for
a slightly general class of theories, described by a prepotential of
the form:
\be \label{em1}
F = - d_{ABC} {X^A X^B X^C\over X^0} - d_A {X^A\over X^0}
\wh A\, ,
\ee
where the indices $A$, $B$, $C$ run from 1 to $N$, and $d_{ABC}$
and $d_A$ are real constants. The corresponding action
describes the low energy effective action of M-theory 
compactified on $S^1\times$
a large volume Calabi-Yau space $\MM$ 
with $N$ four cycles
labeled by the index $A$ ($1\le A\le N$). 
The gauge field $\AAA_\mu^0$ comes from the components
of the metric with one index along $S^1$ and the other index
along a non-compact direction.
On the other hand
the gauge field $\AAA_\mu^A$   arises from
the three form field $C_{MNP}$ with two of the indices 
along the two cycle
of $\MM$ that is dual to the $A$-th four cycle, and the
third  index along a
non-compact direction.  
$d_{ABC}$ are 
the intersection numbers
of the four cycles, and $d_A$ are the second Chern class of the 
four cycles up to a normalization factor\cite{9711053}. 
This class of theories
clearly includes the prepotential \refb{e13} as a special case.
 
First consider a black hole solution for which\footnote{These
black holes have been analyzed in detail in \cite{9711053}. 
Some
recent discussion of these solutions can be found in 
\cite{0603141}.}
\be \label{em2}
p^0 = 0, \qquad q_A = 0, \qquad q_0<0, \qquad
d_{ABC}p^A p^B p^C + 256 d_A p^A >0\, .
\ee
In this case it is easy to show that the following is a solution to
the supersymmetric attractor equations \refb{e11b}-\refb{e11d}:
\ben \label{em3}
&& v_1 = v_2 = 16, \qquad w = 1\, , \nonumber \\
&& x^A = {1\over 8} i p^A, \qquad x^0 = {1\over 8}
\sqrt{d_{ABC}p^A p^B p^C + 256 d_A p^A\over - q_0}\, , 
\nonumber \\
&& e^0 = \sqrt{d_{ABC}p^A p^B p^C + 256 d_A p^A\over - q_0},
\qquad e^A = 0 \quad \hbox{for} \quad A=1,2,\ldots N\, .
\een
The entropy associated with this solution is given by
\be \label{em4}
S_{BH} = 2\pi \sqrt{-q_0 (d_{ABC}p^A p^B p^C 
+ 256 d_A p^A)}
\, .
\ee
These results were first obtained in \cite{9812082}.

For $d_A=0$ the theory, after elimination of the auxiliary
fields, has a $Z_2$ symmetry that allows us to
construct a non-supersymmetric black hole solution from the one
described above by reversing the signs of $q_0$ and 
$e^0$\cite{0511117}. 
In the M-theory description this corresponds to reversing the sign of the
$S^1$ coordinate.
We can
construct the near horizon field configuration associated
with this solution from the one given in \refb{em3} by setting
$d_A=0$, reversing
the sign of $q_0$ and $e^0$ leaving $v_1$, $v_2$ and the $x^I$'s
unchanged, and then finding  $w$ by extremizing 
the entropy
function with respect to this variable. This gives:
\ben \label{em5}
&& v_1 = v_2 = 16,  \nonumber \\
&& x^A = {1\over 8} i p^A, \qquad x^0 = {1\over 8}
\sqrt{d_{ABC}p^A p^B p^C \over  q_0}\, , 
\nonumber \\
&& e^0 = -\sqrt{d_{ABC}p^A p^B p^C \over  q_0},
\qquad e^A = 0 \quad \hbox{for} \quad A=1,2,\ldots N\, , \nonumber
\\ &&
w = {1\over 2}\, ,
\een
for
\be \label{em6}
p^0 = 0, \qquad q_A = 0, \qquad q_0>0, \qquad
d_{ABC}p^A p^B p^C  >0\, .
\ee
It is easy to verify that this configuration extremizes the entropy
function.
The entropy associated with this solution is given by
\be \label{em7}
S_{BH} = 2\pi \sqrt{q_0 (d_{ABC}p^A p^B p^C) } 
\, .
\ee

We shall now calculate corrections to this formula due to
the higher derivative terms proportional to $d_A$. First we note
that for the prepotential given in \refb{em1}, the function $g$
defined through eq.\refb{e9} is invariant under a $Z_2$ symmetry:
\be \label{eg1}
p^A\to p^A, \quad p^0\to -p^0, \quad e^A\to -e^A, \quad
e^0\to e^0, \quad X^A\to -\bar X^A, \quad X^0\to \bar X^0,
\quad w\to \bar w\, .
\ee
{}From the M-theory perspective this corresponds to a change
of sign of the non-compact directions accompanied by
a reversal of the sign of the 3-form field.
Thus for studying the near horizon background associated with
the $p^0=0$, $q_A=0$ black hole we can consider a $Z_2$
invariant configuration:
\be \label{eg2}
p^0=0, \quad e^A=0, \quad x^A=i y^A, \quad x^0=y^0, 
\quad w,\, y^I=\hbox{real}\, .
\ee
In this case the function $g$
defined through eq.\refb{e9} takes the form
\ben \label{egg1}
g &=& {8\over y^0} (v_1 - v_2) 
(d_{ABC} y^A y^B y^C +2\, w^2 d_A y^A) \nonumber \\
&& + {v_2\over v_1}
\left({1\over y^0}\right)^3 (d_{ABC} y^A y^B y^C 
+ 4 w^2 d_A y^A) 
\left(e^0 - {1\over 2} y^0 v_1 w\right)^2 \nonumber \\
&& +\left({1\over y^0}\right)^2
(3 d_{ABC}   y^B y^C + 4 w^2 d_A  )
\left(e^0 - {1\over 2} y^0 v_1 w\right) 
\left(p^A -{1\over 2} y^A v_2 w \right) \nonumber \\ &&
+ 3\, {v_1\over v_2} {1\over y^0} d_{ABC}
\left(p^A -{1\over 2} y^A v_2 w \right) 
\left(p^B -{1\over 2} y^B v_2 w \right)   \, y^C\nonumber \\
&& + {1\over 2}\left({1\over y^0 }\right)^2  
v_2 w (d_{ABC} y^A y^B y^C + 4 w^2 d_A y^A) 
\left(e^0 - {1\over 2} y^0 v_1 w\right) \nonumber \\ &&
- {1\over 2}{1\over y^0 }   v_1 w (3 d_{ABC}   y^B y^C 
+ 4 w^2 d_A  )
\left(p^A -{1\over 2} y^A v_2 w \right) \nonumber \\ &&
-{1\over 4} {1\over y^0} w^2 v_1 v_2
(d_{ABC} y^A y^B y^C + 4 w^2 d_A y^A)
+ 16 {1\over y^0} w^2 
\left(- v_1  - v_2  + {1\over 8}w^2 v_1 v_2 \right) \, d_A y^A
\nonumber \\ &&
+ 128 {1\over y^0} v_1 v_2 (v_1^{-1} - v_2^{-1})^2 \, d_A y^A\, .
\een
This function has a scaling symmetry:
\be \label{eg4}
g(v_1, v_2, w, \lambda^{-1} y^0, \{y^A\},  \lambda^{-1} e^0,
\{p^A\}) =\lambda  \,
g(v_1, v_2, w,   y^0, \{y^A\},    e^0,
\{p^A\})\, ,
\ee
which corresponds to scaling of the $S^1$ coordinate in the M-theory
description.
Now recall that the entropy function  
\be \label{eg5}
\EE = -\pi q_0 e^0 - \pi\, g(v_1, v_2, w,   x^0, \{x^A\},    e^0,
\{p^A\}) \, ,
\ee
has to be extremized with respect
to the variables $v_1$, $v_2$, $e^0$, $y^0$, $y^A$ and $w$.
This can be done by first 
extremizing $g$ with respect to $v_1$, $v_2$,  $w$, $y^0$ 
and $y^A$ and then extremizing the resulting expression
for $\EE$ with respect to $e^0$. Due to the scaling behaviour
given in \refb{eg4}, extremization of $g$ with respect to
$v_1$, $v_2$,  $y^0$ 
and $y^A$ gives a term of the form 
\be \label{eq1}
g=-{K(\{p^A\})\over |e^0|} + {L(\{p^A\})\over e^0}\, ,
\ee
for some functions $K(\{p^A\})$, $L(\{p^A\})$. 
The first term on the right hand side of this equation is invariant
under $e^0\to -e^0$ whereas the second term changes sign under
this transformation. Thus the second term reflects the effect of
parity non-invariant terms in M-theory on the Calabi-Yau manifold
$\MM$.\footnote{Here parity transformation refers to the change of
sign of the $S^1$ coordinate 
without any change in sign of the 
3-form field.}
Substituting \refb{eq1} into
\refb{eg5} gives
\be \label{eg6}
\EE = -\pi q_0 e^0 +\pi\, 
{K(\{p^A\})\over |e^0|} - \pi\, 
{L(\{p^A\})\over e^0}\, .
\ee
Extremizing this with respect to $e^0$ now gives
\ben \label{eg7}
S_{BH}=\EE &=& 2\pi \sqrt{(K(\{p^A\}) - L(\{p^A\})) \, |q_0|}\, ,
\qquad \hbox{for $q_0<0$}\, , \nonumber \\
&=& 2\pi \sqrt{(K(\{p^A\}) + L(\{p^A\})) \, q_0}\, ,
\qquad \hbox{for $q_0>0$}\, ,
\een
assuming that $K(\{p^A\}) \geq |L(\{p^A\})|$.

Eq.\refb{eg7} gives the general form of the entropy in this
theory. 
Comparing \refb{eg7} with \refb{em4} for $d_A=0$ and
\refb{em7} we see that to leading
order $K=d_{ABC} p^A p^B p^C$, $L=0$. 
We shall now calculate the
first non-leading correction to $K$ and $L$, \i.e. corrections of order
$d_A$. Eq.\refb{em4} shows that
\be \label{ekl1}
K(\{p^A\}) - L(\{p^A\}) = d_{ABC}p^A p^B p^C 
+ 256 d_A p^A\, ,
\ee
exactly. To calculate $K+L$ we need to calculate the entropy of the
black hole for $q_0>0$. For this 
we note that since the entropy is the value of
the entropy function $\EE$ at its extremum, an error of order $d_A$
in determining the near horizon background will affect the value of
the entropy function only at quadratic order in $d_A$. Thus to first
order in $d_A$ the computation of the entropy for
$q_0>0$ involves evaluating
the full entropy function in the near horizon background given in
\refb{em5}. This is a straightforward task and yields:
\be \label{em8}
S_{BH} = 
2\pi \sqrt{q_0 (d_{ABC}p^A p^B p^C) } \left( 
1 + {40 d_A p^A \over d_{ABC} p^A p^B p^C}\right) 
+ \OO(d_A d_B)\, .
\ee
This corresponds to
\be \label{eg8}
K(\{p^A\}) + L(\{p^A\})
= d_{ABC} p^A p^B p^C + 80 d_A p^A+ \OO(d_A d_B)\, .
\ee
For the choice $d_{ABC} p^A p^B p^C = p^1 p^2 p^3$ and
$d_A p^A = C p^1$, \refb{em8} agrees with \refb{e28},
\refb{e30} to first
order in $C$.

\sectiono{A Puzzle} \label{s6}

For  theories obtained by dimensional reduction of five dimensional
supersymmetric theories of gravity on a circle,
the
entropy of a class of black holes can be analyzed
using a five dimensional picture\cite{0506176,0508218}.
The black holes discussed in  sections \ref{s4}-\ref{sm}
fall into this class. For these black holes the three dimensional
geometry that includes the compact direction $S^1$, the $AdS_2$
component of the near horizon geometry, and the effect
of the electric field
$e^0$ (regarded as a component of the metric with one index
along $S^1$ and the other index along the time direction) describes
a locally $AdS_3$ space\cite{9712251}, or more precisely the
near horizon geometry of an extremal 
BTZ black hole\cite{9204099}.
Together with the $S^2$ factor
this gives a locally $AdS_3\times S^2$  near horizon 
geometry. The entropy
of such a black hole can then be analyzed using either an
Euclidean action formalism\cite{0506176,0508218,0509148} 
or using
Wald's formalism\cite{9909061,0601228}.  The  answer 
takes the form  
\be \label{ef2}
S_{BH} = 2\, \pi \, \left(\sqrt{c_R h_R\over 6} + 
\sqrt{c_L h_L\over 6}
\right)\, ,
\ee
where $c_R$, $c_L$, $h_R$ and $h_L$ are expressed as functions
of various charges.  
For the black hole solutions described in
section \ref{sm} one finds\cite{0506176,0508218}, 
\ben \label{ef4}
&& h_L = -q_0, \quad h_R = 0, \quad \hbox{for} \quad 
q_0<0\, , \nonumber \\
&& h_L=0, \quad h_R = q_0, \quad \hbox{for} \quad 
q_0>0\, ,
\een
and 
 \be \label{ef3}
c_L = 6 ( d_{ABC}p^A p^B p^C + 256 d_A p^A ), 
\qquad c_R = 6  ( d_{ABC}p^A p^B p^C + 128 d_A p^A)\, .
\ee
On the other hand, \refb{eg7}-\refb{eg8} can be put in the form
given in \refb{ef2}, \refb{ef4} with
\ben \label{ecint}
&& c_L = 6\left( K(\{p^A\}) - L(\{p^A\})\right) =
6 ( d_{ABC}p^A p^B p^C + 256 d_A p^A ), \nonumber \\
&&
c_R = 6\left( K(\{p^A\}) + L(\{p^A\})\right) =
6  ( d_{ABC}p^A p^B p^C + 80 d_A p^A + \OO(d_A d_B))\, .
\een
Comparing \refb{ef3} and \refb{ecint}
we see that our value of $c_L$ agrees with that of \cite{0506176,0508218},
but our value of $c_R$ differs from 
that of \cite{0506176,0508218}.

It is worthwhile reviewing the argument leading to the
computation of $c_R-c_L=12L$ from the five dimensional perspective.
Action \refb{e5}, \refb{e6}
with $F$ given in \refb{em1} has a term proportional to
\be \label{ep1}
d_A \, \int_4  \, Re(X^A/X^0) \, Tr (R\wedge R)\, 
\ee
from the term in the action proportional to $F_{\wh A} \wh C$.
Here $\int_n$ denotes an $n$-dimensional integral.
Since $Re(X^A/X^0)$ can be identified as the component of the
gauge field $\AAA^A$ along $S^1$ in the five dimensional
description, the term \refb{ep1} arises from a term proportional to
\be \label{ep2}
d_A \, \int_5 \,  \AAA^A \wedge Tr (R\wedge R)\, 
= d_A\, \int_5  d\AAA^A \wedge \Omega_3\, ,
\ee
in five dimensions. Here $\Omega_3$ is the gravitational Chern-Simons
term. We can now regard the near horizon geometry
of the black hole solution as a solution in three
dimensional theory, obtained by dimensional reduction of the five
dimensional theory on the $S^2$ factor. Since $p^A$ denotes the
flux of the gauge field strength $\FF^A=d\AAA^A$ through $S^2$,
the three dimensional theory has a term in the action proportional
to
\be \label{ep3}
d_A\, p^A\, \int_3 \,  \Omega_3\, .
\ee
Furthermore this is the only parity non-invariant term in the action
that affects the black hole solution under study. Other possible parity
non-invarint 
terms involving gauge Chern-Simons terms and covariant derivatives
of field strengths and curvature tensor do not contribute in the
background we are considering.  The quantity $L$ can now be computed
in terms of the coefficient of this parity non-invariant term using the
method of \cite{0601228} and gives the answer 
\be \label{ep3a}
L = -64 \, d_A p^A\, .
\ee
This disagrees with the four dimensional result computed from
\refb{ekl1}, \refb{eg8}
\be \label{ep4}
L = - 88 \, d_A p^A +  \OO(d_A d_B)\, .
\ee

The origin of this discrepancy 
is not completely clear to us. Here we discuss
various possibilities. However as indicated in the discussion, we
have been able to rule out most of these possibilities except
the first one.

\begin{enumerate}

\item The analysis of \cite{0506176,0508218} 
applies to the problem at hand only if
the  action and the black hole solution 
that we have used arises, up to a field
redefinition,  from dimensional
reduction of a gauge and general cordinate invariant 
five dimensional theory.
This can be shown to be true in the absence of higher derivative
terms, but has not so far been demonstrated for the theory including
the higher derivative corrections. If the
dimensional reduction of the five dimensional theory produces the
four dimensional theory analyzed here together with an extra set
of terms which are supersymmetric by themselves, the 
discrepancy may be attributed to these missing terms in our
four dimensional action.

\item The analysis of \cite{0506176,0508218} 
uses the Euclidean action formalism as well as the formalism
based on calculation of anomalies in the boundary theory to
compute the entropy of a black hole with near horizon
$AdS_3\times S^2$ geometry. In the absence of the
Chern-Simons term the result for the black hole entropy
agrees with the one computed using Wald's 
formalism\cite{9909061}.
However Wald's formalism cannot be applied directly in
the presence of Chern-Simons terms in the action since the
Lagrangian  density is not manifestly general coordinate invariant.
In contrast our analysis in four dimension is based on Wald's
formalism since the four dimensional Lagrangian density may
be written in a manifestly general coordinate invariant form.
One might wonder if the discrepancy between our result and that
of \cite{0506176,0508218} can be attributed to a difference
between these two formalisms.  This possibility however has been
ruled out in \cite{0601228} 
where the entropy of an extremal
BTZ black hole in the presence of gravitational
Chern-Simons term (and other
higher derivative terms) was computed using Wald's formalism by
regarding this as a two dimensional configuration and
shown to agree with the results of the Euclidean computation.

\item In the analysis of \cite{0506176,0508218} the
quantities $h_R$, $h_L$ are defined as appropriate
conserved charges in the five dimensional theory, while the
quantity $q_0$ is defined as a conserved charge in the four
dimensional theory. The relation 
\refb{ef4} between $h_R$, $h_L$ and the charge
$q_0$ could in principle
be renormalized in the presence of higher derivative terms.
However we have been able to rule out this possibility as
well by regarding the BTZ black hole as a two dimensional
configuration and expressing the entropy of an extremal 
BTZ black hole directly
in terms of the gauge charge in the two 
dimensional theory\cite{0601228}. 
The formula for the entropy takes
the same form as in \refb{ef2} with
$h_R$, $h_L$ replaced by $\pm q_0$
as indicated in \refb{ef4}. After inclusion of the $S^2$ factor
this shows that there is no  renormalization factor between the
conserved charges in five and four dimensions due to the
higher derivative terms.

\item The analysis of \cite{0506176,0508218} relied on
an indirect computation of $c_R+c_L$
based on supersymmetry relations. It is
conceivable that there are subtle effects which affect the various
relations used in \cite{0506176,0508218} 
in arriving at the final formula for  $c_R+c_L$. 
This however does not affect the calculation of
$c_R-c_L=12 L(\{p^A\})$ 
which can be related directly to the gravitational
Chern-Simons term in the five dimensional 
action\cite{0506176,0508218,0601228}. Since our
result for $c_L$ agrees with the five dimensional result while
the result for $c_R$ does not agree, we have a mismatch between
the values of $L(\{p^A\})$ 
calculated using the two descriptions.
This cannot be attributed to a failure of the arguments based
on supersymmetry relations.

\end{enumerate}

In view of the discussion above the only possible explanation seems
to be that the four dimensional action given in \refb{e5}, \refb{e6}
fails to capture some of the terms which come from dimensional 
reduction of a five dimensional supersymmetric 
theory. A probable reason for this is the following.\footnote{This
explanation was offered to us by G.~Lopes Cardoso, B.~de Wit, 
J.~Kappeli and T.~Mohaupt.} The five dimensional supergravity
multiplet, when dimensionally reduced to four dimensions, contains
a gravity multiplet and an additional vector multiplet. Thus if we
add to the five dimensional action  supersymmetrized curvature
squared terms then upon dimensional reduction to four dimensions,
it will contain supersymmetrized curvature squared terms and also
another set of terms which involve supersymmetrization of the four
derivative term involving the additional
vector multiplet fields. In contrast 
the action
used in \cite{9812082,9904005,9906094,9910179,0007195,0009234}
contains the minimal set of terms which are required for 
supersymmetrizing the curvature squared terms. Thus this action
could miss the additional terms involving vector multiplet fields
which would arise from the dimensional reduction of the five dimensional
action.

In view of this it is all the more surprising that for BPS black holes the
result of \cite{9812082,9904005,9906094,9910179,0007195,0009234}
agrees with the one obtined using the five dimensional 
picture\cite{0506176,0508218}. Clearly some additional
non-renormalization theorems which hold only for
supersymmetric black holes are at work here. 
Presumably when the missing terms
are included it will not change the result for supersymmetric
black hole,  but the entropy of the special class of 
non-supersymmetric black holes analyzed in 
sections \ref{s4}-\ref{sm}
will 
agree with the corresponding results derived from the five
dimensional analysis. Once these terms are found, we can calculate
their effect
on the entropy function and  apply
it to calculate the entropy of black holes whose near horizon geometry
do not necessarily have the $AdS_3\times S^2$ form.

\bigskip

\noindent{\bf Acknowledgement}: We would like to thank 
G.~Lopes Cardoso, B.~de Wit, 
J.~Kappeli and T.~Mohaupt for useful correspondence and their
comments on an earlier version of the manuscript. We would
also like to thank Atish Dabholkar, Justin David, 
Rajesh Gopakumar, N~Iizuka, A.~Iqubal, 
Dileep Jatkar,  M.~Shigemori  and Sandip Trivedi 
for useful discussions at various stages of this work.

\end{document}